\documentstyle[12pt,aaspp4]{article}

\begin{document}

\title{Extraordinary Activity in the BL Lac Object OJ~287}
\author{Philip A. Hughes, Hugh D. Aller and Margo F. Aller}
\affil{Astronomy Department, University of Michigan, Ann Arbor, MI 48109-1090;
hughes@astro.lsa.umich.edu, hugh@astro.lsa.umich.edu, margo@astro.lsa.umich.edu}

\begin{abstract}
We use a continuous wavelet transform to analyze more than two decades of
data for the BL~Lac object OJ~287 acquired as part of the UMRAO variability
program. We find clear evidence for a persistent modulation of the total flux
and polarization with period $\sim 1.66$ years, and for another signal that
dominates activity in the 1980s with period $\sim 1.12$ years. The
relationship between these two variations can be understood in terms of a
`shock-in-jet' model, in which the longer time scale periodicity is
associated with an otherwise quiescent jet, and the shorter time scale
activity is associated with the passage of a shock; the different
periodicities of these two components may reflect different internal
conditions of the two flow domains, leading to different wave speeds, or
different contractions of a single underlying periodicity, due to the
different Doppler factors of the two flow components. We suggest that the
modulation arises from a wave driven by some asymmetric disturbance close to
the central engine.  The periodic behavior in polarization exhibits
excursions in $U$ which correspond to a direction $\sim 45^{\circ}$ from the
VLBI jet axis.  This behavior is not explained by the random walk in the
$Q$-$U$ plane which is expected from models in which a pattern of randomly
aligned magnetic field elements propagate across the visible portion of the
flow, and suggests a small amplitude, cyclic variation in the flow direction
in that part of the flow that dominates cm-wavelength emission.
\end{abstract}

\keywords{BL Lacertae objects: individual (OJ~287) --- galaxies: active ---
galaxies: jets --- polarization -- shock waves}

\section{Introduction}
 It is generally accepted that centimeter-waveband emission from AGNs is
associated with a jet of synchrotron plasma, the accretion structure and
immediate environment of the central supermassive black hole contributing
broadband emission from the infrared to $\gamma$-rays (e.g., Bregman
\markcite{REF09} 1994). A number of processes may be responsible for temporal
variations in the radio flux: the fueling (accretion) rate may change with
time, leading to a long-term change in the jet power; the accretion disk may
exhibit instability, influencing the extraction of energy in the form of a
collimated outflow; the outflow itself may be Kelvin-Helmholtz unstable,
leading to propagating internal structures (e.g. shocks); the outflow may
interact with ambient inhomogeneities, which also may produce disturbances to
the body of the flow (Wiita \markcite{REF40} 1991; Birkinshaw
\markcite{REF06} 1991; Icke \markcite{REF20} 1991). However, while there can
be characteristic time scales associated with all such process, there is no a
priori reason to believe that such variations will be {\it periodic}.
Nevertheless, a detailed understanding of blazars demands a detailed
description  of the temporal variations, which for completeness requires a
search for periodicity.

 OJ~287 is a BL~Lac object with $z=0.306$ (Miller, French, \& Hawley
\markcite{REF29} 1978; Sitko \& Junkkarinen \markcite{REF35} 1985), whose
behavior differs in a number of intriguing ways from that of other blazars.
Large amplitude, short time scale variations (Sillanpa\"a et al.
\markcite{REF34} 1994; Kidger \& Gonz\'alez-P\'erez \markcite{REF25} 1994;
Valtaoja, Ter\"asranta, \& Tornikoski \markcite{REF38} 1994; Aller, Aller, \&
Hughes \markcite{REF02} 1994) are seen in the light curves from the radio to
optical wavebands. In the optical waveband, OJ~287 displays activity down to
a time scale $\sim 10$ minutes (Boltwood \markcite{REF08} 1996). Numerous
periodicities have been claimed, the strongest evidence being for a period
$\sim 12$ yr evident in optical data that span more than one century
(Sillanpa\"a et al. \markcite{REF33} 1988).  This period has been interpreted
by Valtonen and coworkers (e.g., Valtonen \& Lehto \markcite{REF39} 1997) as
being due to the passage of the secondary black hole of a binary system
through the accretion disk of the primary black hole.  Interpreting VLB
polarization observations in the context of a `shock-in-jet model', Cawthorne
\& Wardle \markcite{REF11} (1988) suggests that the radio jet is aligned very
close to the critical cone, $\theta_c=\sin^{-1}\left(1/\gamma\right)$, in
contrast with viewing angles of tens of degrees suggested by the modelling of
events in some blazars (Hughes, Aller, \& Aller \markcite{REF17} 1989,
\markcite{REF18} 1991; Mutel et al. \markcite{REF30} 1990; Carrara et al.
\markcite{REF10} 1993).  This suggests that the flow should have a Doppler
factor significantly in excess of unity, with a concomitant reduction in
observed time scales -- making the source uniquely suited for an analysis of
periodicity, because of the many cycles observed. Indeed, Katz
\markcite{REF24} (1997) has proposed a model based on a precessing disk, that
would demand an extremely high Lorentz factor ($\sim 50$), and thus a high
Doppler factor if the flow is seen from close to, or within, the critical
cone.

 Motivated by the preceding arguments, and by the intriguing claims of
periodicity for the optical light curve of OJ~287, we performed a number of
spectral analyses of centimeter waveband data for those sources well-enough
observed to justify a structure function analysis (Hughes, Aller, \& Aller
\markcite{REF19} 1992). Both Deeming \markcite{REF12} (1975) and Scargle
\markcite{REF32} (1982) periodograms were constructed, together with smoothed
power spectra using both standard tapering and smoothing techniques (Jenkins
\& Watts \markcite{REF21} 1968), and a maximum entropy approach (Haykin \&
Kesler \markcite{REF15} 1979; Ulrych \& Ooe \markcite{REF37} 1979). In
general, each source exhibited a complex distribution of power with
frequency, due in part to the fact that most activity is aperiodic, and in
part to the observing window.  In only one source -- OJ~287 -- was there some
evidence for periodicity; in that case the Scargle false-alarm probability
gave us some confidence in a variation with time scale $\sim 1.6$ yr (Aller,
Aller, Hughes, \& Latimer \markcite{REF03} 1992).

 Although a powerful technique in the right context, the fundamental
limitation of Fourier methods is that they do not preserve the temporal
locality of a signal.  Power associated with the data's window will appear at
low frequencies, possibly near to the frequency of a sought periodic signal,
while outbursts with a character that is certainly not periodic will place
power across the spectrum. Identifying periodicity that is well-hidden by
other variations requires assessing the significance of one, among many,
peaks in the power spectrum. A method of analysis that circumvents this
problem is continuous wavelet analysis, which we discuss in detail in \S~3.
For a one-dimensional data set such as a time series, a wavelet analysis
amounts to quantifying the behavior of the signal on different temporal
scales, as a function of time. This is achieved by convolving the signal with
a localized wave-packet, as the packet is translated along the series, for a
number of `dilations' of the wave-packet, i.e., for progressively broader
wave-packets, sensitive to progressively longer time scale behavior.  Thus an
initially narrow wave-form, sensitive to high frequency structure in the
signal, is progressively stretched, or dilated, in time, making it sensitive
to lower frequency structure. We are currently exploring the behavior
exhibited by various sources in the UMRAO database using continuous wavelets,
and here report results for OJ~287, having chosen this source as the first to
study in light of its unique character.

\section{Centimeter Waveband Data}
 Figure~1 shows the UMRAO data for OJ~287 averaged in 30-day periods to
reduce the crowding of data points.  Observations were made at approximately
weekly intervals at each frequency starting in 1978, 1971 and 1974 at 4.8,
8.0 and 14.5~GHz respectively. There is a systematic gap in the data at
yearly intervals (of approximately 30-day duration) to avoid solar
interference when the sun is too close to the source. There are other gaps of
varying length in the data, due to bad weather and equipment malfunctions,
but these are randomly placed throughout the observing period. Selected HII
regions, Galactic supernova remnants and radio galaxies were observed each
day to determine the effective collecting area of the telescope and to verify
the stability of the instrumental polarization.  The observing technique and
data reduction procedures are discussed more fully by Aller et al.
\markcite{REF04} (1985, \markcite{REF05} 1997). The wavelet analysis that
follows uses the unaveraged data: 583 points at 4.8~GHz, 1060 points at
8.0~GHz and 828 points at 14.5~GHz.

\section{The Continuous Wavelet Transform}
 The fundamental concept behind continuous wavelet analysis is to convolve
the `signal' under study with a translated and dilated `mother' wavelet, in
order to map out in translation-dilation space the power in a signal at a
particular time, on a particular scale. The great advantage of this over
Fourier techniques is the preservation of temporal locality: a gap in the
time series will be evident along the corresponding line in transform space,
and events that are distinct in the signal will have distinct counterparts in
transform space. A corollary of this is that there is a redundancy in the
wavelet transform of a periodic signal -- a set of peaks and troughs in the
real part of the transform, that `march' across transform space at fixed
dilation -- that makes the detection of periodicity very easy.  A number of
reviews and books addressing this subject have appeared since the early
1990s; we follow here a review by Farge \markcite{REF13} (1992) on the
application of wavelet analysis to the study of turbulence.

The Morlet wavelet has been extensively used in a number of fields of study,
and it has the advantage of being both complex and progressive. Being
complex, the real part of the transform exhibits an oscillatory behavior
corresponding to periodicity of the signal being analyzed, thus highlighting
such periodic behavior, while we can construct the modulus, which because of
the phase shift between the real and imaginary parts of the wavelet
coefficients, provides a smooth estimation of the power in the signal being
analyzed, and its temporal persistence. The frequency of a periodic signal
may be read from the phase plot, which also provides a sensitive diagnostic
of both changes in characteristic behavior of the signal and sharp pulses.
Being progressive, i.e., having zero power at negative frequency, the wavelet
is optimal for the analysis of causal signals, because it does not admit an
interference between the past and future parts of the signal at any instant
in the time series. The wavelet is characterized by a frequency
$\omega_{\psi}$:
\begin{equation}
\psi_{\rm Morlet}=e^{-i\omega_{\psi}t}e^{-|t^2|/2},
\end{equation}
and has Fourier transform
\begin{equation}
\hat\psi\left(\omega\right)=\left(2\pi\right)^{-1}\int_{ {\bf R}}\psi
\left( {\bf t}\right)e^{-i\omega t} dt,
\end{equation}
given by
\begin{equation}
\hat\psi\left(\omega\right)=\cases{\raise 1.0ex \hbox{$\left(2\pi\right)^
{-1/2}e^{-\left(\omega- \omega_{\psi}\right)^2/2}$} &if $\omega> 0$;\cr 0
&if $\omega \leq 0$.\cr}
\end{equation}
To provide a good resolution of temporal structures we want to choose
$\omega_{\psi}$ to be small (the higher the frequency of the analyzing
wavelet, the greater the extent to which the wavepacket is a broad
oscillatory structure, which `delocalizes' structure in the signal, as does
Fourier transformation -- just what we are trying to avoid); however, the
mother wavelet must satisfy an admissibility condition
\begin{equation}
C_{\psi}=\left(2\pi\right)\int_{{\bf R}}|\hat\psi\left({\bf \omega}\right)|^2
{d\omega\over \omega} < \infty.
\end{equation}
Equation (4) is equivalent to requiring a zero mean:
\begin{equation}
\int_{ {\bf R}}\psi\left(t\right)dt=0.
\end{equation}
This condition is not formally satisfied by the wavelet of equation (1) for
any value of $\omega_{\psi}$; however if $\omega_{\psi}$ is large enough, the
admissibility condition is satisfied to the single precision machine accuracy
with which wavelet coefficients are computed. A value of $\omega_{\psi}=5$
reconciles the conflicting requirements of resolution and admissibility; we
have performed the analysis described below with different values of
$\omega_{\psi}$, with no discernible change in the transform coefficients.

 Having selected a mother wavelet, the wavelet coefficients are found by
constructing a set of translated ($t'$) and dilated ($l$) wavelets
\begin{equation}
\psi_{lt'}\left(t\right)=l^{-1/2}\psi\left[{t-t'\over l}\right], \ \  l\in
 {\bf R}^+,\ \ t\in {\bf R},
\end{equation}
and convolving with the signal:
\begin{equation}
\tilde f\left(l,t'\right)=\int_{ {\bf R}} f\left(t\right) \psi_{lt'}^*
\left(t\right)dt.
\end{equation}
We have chosen to normalize so that all wavelets have the same $L^2$ norm so
that wavelet coefficients correspond to energy densities. The discrete nature
of the data, of course, means that the above integrals have to be
approximated as sums, and we evaluate $\tilde f\left(l,t'\right)$ on a mesh
of values $\left(l_i,t_i'\right)$, fine enough to display all the structure
associated with the input signal. The data are well-sampled with an {\it
approximately} constant density of sampling points on time scales
significantly shorter than that of the spectral features discussed here, and
contain few gaps. To facilitate the computations, we interpolate the data
onto a uniform sampling grid with scale $\Delta t$, and compute the wavelet
coefficients up to a dilation $l_{\rm max}$ corresponding to a time scale of
5 years.  The interpolation has no significant influence on the signal for
structures with scale $>>\Delta t$. The convolution that determines the
wavelet coefficients $\tilde f\left(l,t'\right)$ requires data within a
domain of influence $t'-l\lesssim t \lesssim t'+l$, and so the larger the
dilation, $l$, the further does the effect of the ends of the time series
extend. The conservative choice of a 5 year limit ensures good coverage of
transform space unpolluted by edge effects. In the transform space plots
shown below we have masked out the wavelet coefficients within $1.5l$ of the
boundaries, the domain in which the coefficients cannot be reliably
computed.  Note that these plots are linear in translation, and logarithmic
in dilation.

 As examples of this technique, Figures~2a and 2b show the transforms of
Gaussian white noise and a sinusoidal signal in which the frequency halves at
the mid-point of the time series respectively. A short period of `coherent'
activity in the white noise signal leads to structure in the real part of the
transform coefficients that might be mistaken for periodicity.  Examination
of the modulus, however, shows that this feature is transitory.

The admissibility condition guarantees that the wavelet coefficients may
be used to reconstruct the original signal:
\begin{equation}
f\left(t\right)=C_{\psi}^{-1}\int_{{\bf R}}\int_{{\bf R}^+} \tilde
f\left(l,t'\right) \psi_{lt'}\left(t\right){dl\,dt'\over l^2},
\end{equation}
and the fact that nearby wavelet coefficients are correlated corresponds to a
redundancy in the set of coefficients $\tilde f\left(l,t'\right)$ which means
that a wavelet different from the analyzing wavelet may be used in the
reconstruction; in fact, a delta distribution may be used, leading to the simple
reconstruction formula
\begin{equation}
f\left(t\right)= C_{\delta}^{-1} \int_{{\bf R}^+}\tilde f\left(l, t\right)
{dl\over l^{1+\left(1/2\right)}},
\end{equation}
where
\begin{equation}
C_{\delta}=\left(2\pi\right)^{1/2}\int_{ {\bf R}}\hat\psi\left(\omega\right)
{d\omega\over \omega}.
\end{equation}
We cannot reconstruct the original signal exactly because we compute the
transform coefficients for only a finite set of translations and dilations.
In particular, the upper bound on $l$ implies that the wavelet coefficients
contain no information about long-term trends (beyond about 5 yr), and
effectively subtracts a local mean. The main purpose of reconstruction is to
display that part of the signal within limited ranges of $l$; we can assess
the validity of that by comparing the reconstruction that uses all available
dilations with the original signal. Use of a discrete transform in which the
analyzing wavelets form a complete orthogonal basis (Farge \markcite{REF13}
1992) would provide the ability to invert exactly, but the continuous
transform used here provides a much more useful method for illuminating the
nature of the signal variations, and in particular highlighting
periodicities.

\section{Analysis of the Total Flux Variations}

\subsection{The Transform Coefficients}
 Figure~3 shows the Morlet transform for the 14.5~GHz flux data for OJ~287.
Some power is evident over the whole range of time scales explorable within
the time series, but the most obvious feature is the signature of periodicity
at low frequency, in the form of a pattern of peaks and troughs that crosses
the transform plane at fixed dilation. This pattern is modified during the
1980s, by the appearance of shorter period activity, but resumes after the
latter activity has subsided. Evidently this shorter period activity results
from a coexisting emission component, because both signals may be seen
distinctly at some epochs, e.g., 1983, and hints of the shorter period
activity persist into the 1990s. The transforms of the 8.0 and 4.8~GHz data
are similar to the one shown. However, the interpretation of the 4.8~GHz
transform is somewhat less clear-cut because of structure in transform space
on the longest time scales accessible to study; this presumably results from
the fact that at the lowest observing frequency we are including
contributions from larger scale flows (Marscher \markcite{REF27} 1993). For
comparison, Figures~4a and 4b show the transforms of two sources observed in
the Michigan program that do not exhibit the periodic type of activity seen
in OJ~287.  A short period of `coherent' activity in the major outburst seen
in BL~Lac in the early 1980s leads to structure in the real part of the
transform coefficients that might be mistaken for periodicity.  Examination
of the modulus, however, shows that this feature is transitory, and quite
different from the persistent modulation seen in OJ~287. There is a hint of
periodic behavior in NRAO~140, seen in the real part of the transform
coefficients at the longest dilation, between the mid-1980s and mid-1990s,
but on time scales too long to be explored with the current data.

To assess the probability that the transform of OJ~287 is significantly
different from that of Gaussian white noise, we have computed a
$\chi^2$-statistic based on a pixel-by-pixel comparison of the two
transforms. As we are concerned with a comparison of the large amplitude
structures evident at high dilations, and also wish to judge the validity of
a model presented in \S 4.2 -- which makes no attempt to fit the low level
structure of the upper half transform plane, we have used the standard
deviation of the transform modulus in the upper half plane as an estimate of
an `error' in the OJ~287 transform. We find a $\chi^2$ probability of $\sim
100\%$ that the model discussed below fits the data, but a probability of
$\la 10^{-5}\%$ (i.e., zero to machine accuracy) that Gaussian white noise
can model the data, for a number of realizations of a noise-based signal.

 The frequency of the modulation is not readily extracted from the transform
plot, because the analyzing wavelet picks out a scale $\zeta l$ where, for
the Morlet wavelet adopted here, $\zeta\sim 1.2$, but is not a precisely
defined number. From the location in transform space of the persistent
pattern, the period is $\sim 1.6$ yr. We may use a discrete form of the
inversion formula given in the last section to reconstruct that part of the
signal corresponding to a particular range of dilations, and have done that
for two juxtaposed bands in this case. Specifically, we have made cuts just
longward in period of the persistent modulation, and just shortward in period
of the shorter period signal that peaks in 1985. We have divided that band at
a dilation that optimizes the separation of the two periodic components.
Figure~5 shows the reconstructed signal in each of the two bands.  There is
some `leakage' between bands near to the center of the time series, but the
periodic character of the variations is evident. Fitting a sinusoid to the
inner-decade in each case yields periods of $1.66$ yr and $1.12$ yr.  (In the
cosmological comoving frame of the source, these periods are $1.27$ yr and
$0.86$ yr respectively.) The fact that the longer period activity is a stable
modulation of the underlying flow is evident from its phase coherence.
Figure~6 shows this oscillation, superposed on which is a sinusoid of period
$1.66$ yr, and with the second half the reconstructed variation folded back
on itself.  Figure~7 shows the sum of the reconstructed signals and a tapered
boxcar average of the original data. (Recall that the transform removes a
local average.) Although the earlier periodogram analysis failed to provide
clear-cut evidence for periodicity because of a smearing of power between
these two signals, clearly a substantial portion of the source variation is
accounted for by the two periodic components.

 We have made a similar reconstruction at 4.8~GHz, and for both of the two
bands of dilation have computed the spectral index, $\alpha$, defined in the
sense $S\left(\nu\right)\propto \nu^{-\alpha}$. (As this requires us to take
the ratio of fluxes, which obviously is not meaningful for quantities that
oscillate about a zero mean, for each of the two dilation bands we have
computed $\alpha$ from the fluctuating portion of the signal plus a tapered
boxcar mean.) Figure~8 displays the varying spectral indices, the bounds of
which remain within values typical for flat spectrum sources (Aller, Aller,
\& Hughes \markcite{REF01} 1992), with the spectra being most self-absorbed
($\alpha<0$) when the flux is at its highest. This is of course consistent
both with an increase in flux being associated with an increase in intrinsic
opacity, or with an increase in flux being associated with an increased
Doppler boost with concomitant decrease in rest frame frequency -- and so an
increase in opacity -- corresponding to a given observational frequency. Thus
the spectral behavior of the source components does not allow us to
discriminate between models for the periodic variation, but is consistent
with the obvious candidates: a modulation of the flow density or direction.

To look for corresponding variations in the optical data, we have used
Rosemary Hill Observatory data kindly supplied by S. Clements. We have
performed a wavelet analysis of optical waveband variations from $\sim 1970$
to $\sim 1990$, a time period comparable to that covered by the radio data.
Some structure is seen in the transform coefficients at the longest time
scales, but no periodicity is evident, in particular on scales of order $1$
yr. Both the radio data and the available Rosemary Hill Observatory optical
data time series are, of course, too short to probe the $12$ yr time scale
evident in the historical optical data.

\subsection{Interpretation: A Shock-In-Jet Model}
 We believe that an effective way to account for the behavior of the
transform is with a `shock in jet' model (Blandford \& K\"onigl
\markcite{REF07} 1979; Marscher, Gear, \& Travis \markcite{REF28} 1992), in
which a region of enhanced emission (the down stream flow of the shock)
travels along an otherwise quiescent flow, bounded by the $\tau=1$ surface
(behind which the flow is optically thick, and the radiated flux is
negligible) and a region of severe adiabatic energy loss, which also
contributes negligible flux. Both effects depend so sensitively on position
within the source, that to a good approximation the quiescent flow can be
regarded as `windowed' within axial extent $z_1\leq z\leq z_2 $. The
propagating shock displaces {\it part} of the quiescent jet emission, as
required by the transform behavior.

 Consider a quiescent jet with specific emissivity a function of distance
$z$ along the flow:
\begin{equation}
d\varepsilon_j = z^{-\chi}dz; \ \ \ z_1\leq z\leq z_2,
\end{equation}
along which a shock propagates with emissivity enhancement $f$:
\begin{equation}
d\varepsilon_s = f \times z^{-\chi}dz; \ \ \ z_A\left(t\right)\leq z \leq
z_B\left(t\right).
\end{equation}
Within any segment of the flow the total emission is
\begin{eqnarray}
\varepsilon^i & = & \int^{z_{{\rm hi}}}_{z_{{\rm lo}}} \left[f \right]
 z^{-\chi}dz \nonumber \\  & = & \left[f \right]{z^{-\chi+1} \over -\chi+1}
 \Biggl| ^{z_{{\rm hi}}}_{z_{{\rm lo}}}
\end{eqnarray}
where the factor $f$ appears if this segment is within a shock. For example,
when the shock is wholly within the windowed portion of the flow, we compute
the total emission from the three segments: $z_1\leq z\leq z_A $, $z_A\leq
z\leq z_B $ and $z_B\leq z\leq z_2 $. Now suppose that the emission is
sinusoidally modulated, with slightly different frequencies in the quiescent
and shocked portions. As the transform subtracts a mean, we need consider
only the fluctuating portion of the signal, and observe
\begin{equation}
\varepsilon_{\rm tot} = \sum_i \varepsilon_j^i\left(t\right) \sin\left (2\pi
\nu_j t\right) + \varepsilon_s^i\left(t\right) \sin\left (2\pi \nu_st\right).
\end{equation}

 We adopt a simple model for the evolution:
\begin{equation}
z_A = z_0\left(t\right)-{\Delta \over 2}; \ \ \ z_B = z_0\left(t\right)+
    {\Delta \over 2}; \ \ \ \Delta={1\over 2} \left(z_2-z_1\right)
\end{equation}
where
\begin{equation}
z_0\left(t\right)= \left[3\left(z_2-z_1\right)\right]\left[{t\over t_{\rm 
     max}}\right]+\left(2z_1-z_2\right).
\end{equation}
This describes a component propagating with constant speed, and, with
appropriate choice of $z_1$ and $z_2$, of such a size as to provide
`activity' with the observed duration as a fraction of the whole time
series.  From the wavelet transform we know that $\nu_j = 0.60\,{\rm
yr}^{-1}$ and $\nu_s = 0.89\,{\rm yr}^{-1}$. It is possible that this
difference originates from the different Doppler factors associated with
these flow segments, there being an underlying frequency $\nu_0$ Doppler
contracted to $\nu_0/{\cal D}_j$ in the case of the quiescent jet, and
contracted to $\nu_0/{\cal D}_s$ in the case of the shock. Another
possibility is that different physical conditions in jet and shock lead to
different natural frequencies for the development of oscillations. Values
of other parameters that provide a good fit to the data are $z_1=1$ and
$z_2=10$ and $\chi=1.5$, $f=3.5$.

 Figure~9 shows the Morlet transform of the model signal, which displays the
main characteristics of the data transform.  The model assumes a single event
in the 1980s, whereas that activity might be a due to any number of closely
spaced structures with enhanced emissivity. The key point is that any
activity with the general character adopted for the model can account for the
evolution seen in transform space, and so provides support for a persistent
modulation of an underlying flow.

 The origin of this persistent modulation is unknown, but seems unlikely to
be related simply to a helical bulk flow (e.g., Steffen et al.
\markcite{REF36} 1995; Vicente, Charlot, \& Sol \markcite{REF42} 1996),
perhaps induced by precession of the central engine:  the only way that
helicity alone could produce a modulation of the emissivity is if a
propagating structure has its instantaneous velocity vector directed
periodically towards the observer as the structure traverses consecutive
turns of the helix; but that would require a single event to persist for many
turns, for several decades of time. The inevitable rotation of the outflow
resulting from an accretion structure wind or magnetic torque (Wiita
\markcite{REF40} 1991) is not sufficient to explain the observations, as this
does not break cylindrical symmetry.  It is more likely that a disturbance
drives a wave along the jet, so that the velocity vector of any comoving
volume of jet material  describes a helical trajectory due to its combined
forward (axial) motion and a cyclical, transverse (precessional-like)
motion.  The velocity vector associated with the fixed, visible length of the
flow -- `windowed' by opacity near to the core, and by severe adiabatic
energy loss away from the core -- will shift cyclically, and persistently, as
does the thread of a screw at the surface through which it is being driven.
This may be a manifestation of the Kelvin-Helmholtz instability:  in
principle, the surface helical mode just introduces a cyclic transverse
motion with the amplitude related to the angular frequency and the amplitude
of the helical twist; body and higher order modes will be more complicated
but all imply some precession of the velocity vector (P. Hardee, private
communication).  The modulation has a periodicity similar, but not identical,
to the $1.2$ yr value discussed by Katz \markcite{REF24} (1997) as due to a
`nodding' of a precessing structure, but it is not obvious how such a motion
would influence a jet.

Recently published VLBI images of OJ~287 (Gabuzda \& Cawthorne
\markcite{REF14} 1996) show components typically from $1$ to $7$ pc from the
core (assuming $q_0=0.5$ and $H_0=75\,{\rm km}\,{\rm s}^{-1}\,{\rm
Mpc}^{-1}$). If we take $5$ pc as a typical knot distance, and assume a flow
opening angle of $1^{\circ}$ (Marscher \markcite{REF26} 1987; Muxlow \&
Garrington \markcite{REF31} 1991), the radius of the flow at that point is
$\sim 0.05$ pc.  The flow cannot rotate faster than the internal sound speed,
and if that is $c/\sqrt{3}$ -- assuming a relativistic plasma -- the rotation
time will be $\geq 1.5$ yr. It seems more plausible that a rotation of the
flow would occur more slowly, but the bulk of the flux will probably be
associated with the vicinity of the innermost knot, reducing the
characteristic length scale, and permitting a rotation time as short as
years. If this is, indeed, the explanation for the observed modulation, the
latter is due to a nonrelativistic pattern, and would not be Doppler
contracted. In that case, the different periodicities associated with jet and
shock would reflect different internal states, and thus different phase
speeds for wave propagation.

\subsection{Constraints on Angle and Speed of Flow}
 We have noted that the different frequencies evident in the wavelet
transform might reflect a single intrinsic frequency of modulation, Doppler
contracted by different amounts in different segments of the flow. If that
is so, we can use the ratio of frequencies to constrain aspects of the source
dynamics; conversely, we can ask whether in the context of a `shock-in-jet'
model it is plausible that the observations are explained in this way. For a
flow of speed $\beta$ and Lorentz factor $\gamma$, viewed at angle $\theta$
to the flow axis, the Doppler factor is
\begin{equation}
{\cal D}=\frac{1}{\gamma\left(1-\beta\cos\theta\right)}.
\end{equation}
Time scales are contracted by this amount, so that for flows modulated
at a period $P_0$, but contracted to periods $P_1$ or $P_2$
\begin{equation}
{\cal R} = \frac{{\cal D}_2}{{\cal D}_1} = \frac{P_1}{P_2}.
\end{equation}
If $P_1$ and $P_2$ are the periods of the persistent, jet flow, and the
enhanced, perhaps shocked flow respectively, ${\cal R}=1.48$.

 Suppose first that the flow speed is unchanged, and that the change in
Doppler factor arises solely from a change in flow orientation, $\theta
\rightarrow \theta+\Delta\theta$. If $|\Delta\theta|<<1$ then using the
apparent speed
\begin{equation}
\beta_a=\frac{\beta\sin\theta}{1-\beta\cos\theta},
\end{equation}
we may write
\begin{equation}
1+\beta_a\Delta\theta={\cal R}^{-1},
\end{equation}
where we interpret $\Delta\theta$ as the change from the initial flow to
enhanced flow phase. With the caveat that knots may represent a pattern
speed, and not the speed of the underlying flow, we can use the observed
superluminal motion of OJ~287 (Gabuzda \& Cawthorne \markcite{REF14} 1996;
Gabuzda, Wardle, \& Roberts \markcite{REF41} 1989) to estimate that
$\beta_a\sim 4.4$, for the cosmological parameters adopted (see above). This
value was measured in the early 1980s, and is therefore indeed appropriate to
the initial flow. Moreover, there is only a modest dispersion in the apparent
speeds measured for the three components visible in the 1980s, justifying the
use of a single value throughout the time of interest. From the values of
$\beta_a$ and ${\cal R}$ we have $\Delta\theta =-4.2^{\circ}$.  Thus, a
fairly modest change in flow direction is sufficient to explain the change in
Doppler factor necessary to account for the change in observed periodicity.
For a flat spectrum source in which the enhanced emission fills the
observable `window' of the flow -- to a first approximation -- the observed
flux will be increased by the increased Doppler boost, ${\cal B}\approx {\cal
D}^2$, which implies a flux increase by a factor of $\sim 2.2$; this could
certainly account for much of the brightening observed in the 1980s.  The
primary reason for discounting this scenario is that as the shorter and
longer period signals coexist at some epochs, and as hints of the shorter
period signal are evident at late epochs, we would require that only a
portion of the flow reorients itself by $\sim 4^{\circ}$, and then most but
not all of that flow segment relaxes back to its original configuration.
Furthermore, the scenario is implausible because we would expect a continuous
range of viewing angles for a curved flow, not the essentially bimodal
distribution necessary to produce two distinct periods.

The obvious alternative scenario is that the flow speed changes: $\beta
\rightarrow \beta+\Delta\beta$, $\gamma \rightarrow \gamma+\Delta\gamma$,
where $\Delta\gamma=\beta\gamma^3\Delta\beta$. Consider such a change, in
which the flow orientation is fixed. As before, we can write
\begin{equation}
1-\left(\gamma{\cal D}-\gamma^2\right)\left(\frac{\Delta\beta}{\beta}\right)-
\left(\gamma{\cal D}-1\right)\gamma^2\beta^2\left(\frac{\Delta\beta}{\beta}
\right)^2={\cal R}^{-1}.
\end{equation}
If we are viewing close to the critical cone of the flow, then ${\cal D}
\sim \gamma\sim \beta_a$, so that the second term of the previous expression
is negligible, and
\begin{equation}
\Delta\beta=\left[\frac{1-{\cal R}^{-1}}{\left(\beta_a^2-1\right)\beta_a^2}
\right]^{1/2}.
\end{equation}
Using our previous parameter values, we find that $\Delta\beta=0.03$, while
the corresponding (arithmetic) change in Lorentz factor is
$\Delta\gamma=2.6$.  The inferred $\Delta\beta/\beta$ is barely small enough
to justify the simple linearization used above, but the numbers are certainly
illustrative of the change required.  A change of $\gamma$ from $4.4$ to
$\sim 7$ requires a Lorentz transformation by only $\gamma\sim 1.1$,
consistent with the result that much source activity can be explained by weak
shocks (Jones \markcite{REF22} 1988).

\section{Analysis of the Polarized Flux Variations}

\subsection{The Transform Coefficients}
 A wavelet analysis identical to that applied to the total flux has been used
to explore variations in the polarized flux. We have made separate studies of
the $Q$ and $U$ Stokes parameters, and Figure~10 displays the signal and
transform coefficients for 14.5~GHz $U$ data for OJ~287. The same persistent
signal as seen in the total flux is evident here. The influence of mid-1980s
activity is barely apparent, which may be explained by the `enhanced flux'
component being more opaque, and contributing little to the polarized flux;
however, as we discuss below, there is evidence for a change in character of
the polarized fluctuations at this time. A longer period signal is also
apparent -- just above the lower edge of the plot of the real part of the
transform, during the 1980s -- which may be more evident in the polarized
flux than in the total flux, because optically thin, and thus probably more
extended, structures will tend to dominate the polarization. This possible
longer period signal cannot be explored with a time series of several decades
in length, and we do not discuss it further.  The transform of $Q$ also
displays the longer period signal, but the persistent $1.66$ yr signal is
almost absent, indicating that the fluctuations are almost entirely in $U$.
The transforms at 8.0 and 4.8~GHz are very similar to that at 14.5~GHz.

Examination of $Q$-$U$ plots for the other two observing frequencies shows a
similar pattern of behavior, but with different orientation to the axis of
variations. We have explored this by computing transforms for both $Q$ and
$U$, with the $Q$-$U$ plane systematically rotated, in each case looking to
see what rotation maximizes the variations in $U$. As a measure of the
variations, we have integrated the reconstructed fluctuating signal over all
times (translations). At 14.5~GHz a rotation of $-12^{\circ}$ was needed to
maximize $U$, while at 8.0 and 4.8~GHz the angles were found to be
$0^{\circ}$ and $30^{\circ}$ respectively. This serves to quantify the degree
to which variations relate to the VLBI jet directions, but also suggests a
systematic variation with frequency that may be due either to the presence
of a Faraday medium, or to the fact that at different observing frequencies,
observations are sensitive to different parts of the flow, i.e., the $\tau=1$
surface moves inwards at higher frequency.

 We have reconstructed the periodic part of the signal, using the range of
dilations adopted for the total flux, for both $Q$ and $U$, and Figure~11
displays the evolution of this signal at 14.5~GHz for the longer period of
the two bands adopted earlier. Note that as the transform removes the mean,
the points are placed about the origin of the plane. To provide a more
meaningful indication of $Q$-$U$ plane evolution, we have computed a tapered
boxcar average of the original signal, and in Figure~12 show the sum of this
and the fluctuating part. The evolution is most easily seen when displayed
dynamically (an animation is available at the primary author's WWW site:
http://www.astro.lsa.umich.edu/users/hughes/), and has the character of
smooth `orbits' about the origin. Looking only at the modulated component
(Figure~11) we see that the source oscillates in the $Q$-$U$ plane, initially
along an axis approximately coincident with the direction of the VLBI jet
($-110^{\circ}$; Gabuzda \& Cawthorne \markcite{REF14} 1996), but thereafter
along an axis orthogonal to that, until the variations fall back to their
original character towards the mid-1990s. As will be demonstrated in the next
section, the smoothness of variation is an artifact of the filtering: by
reconstructing the signal from a limited range of dilations, the short period
fluctuations that normally endow $Q$-$U$ plots with a highly irregular
character have been removed.  Nevertheless, the analysis does highlight that
during the activity of the 1980s the polarized flux varied almost entirely in
$U$, which given the jet orientation, corresponds to `excursions' of the
electric vector about $45^{\circ}$ to the flow axis.

 Examination of Figure~12 shows that at early epochs, the polarized flux was
primarily in $-Q$, corresponding to an electric vector orthogonal to the VLBI
flow, while during the more active phase that follows, a significantly
fluctuating electric vector more closely aligned with the flow occurs. This
is qualitatively consistent with a picture in which an axial magnetic field
gives way to a transverse field during activity -- as envisaged by the
standard `shock-in-jet' models (Hughes, Aller, \& Aller \markcite{REF16}
1985, \markcite{REF17} 1989) -- but with the effective field orientation
modulated by a corkscrew-like wave of the internal velocity vector of the
flow as discussed above, with further complexity perhaps added by the effects
of relativistic aberration.

The relationship between the variations in $I$ and those in $U$ is somewhat
complicated, but the trends may be exposed by comparing and combining the
reconstructed signals from both the low and high frequency dilation bands.
Figure~13 shows the cross-correlation function of $I$ and $U$ variations in
the low frequency band. The correlation function behaves in a manner very
similar to that expected from two sinusoidal signals, which are out of phase
by $\pi$, leading to a strong anticorrelation at zero lag. Such behavior is
evident from Figure~14, which shows the product of these two signals. The
broken horizontal line near to the top of the plot marks where the two
signals have the same sign. Although there are epochs that run counter to
this trend, as a general rule the $I$ and $U$ signals are anticorrelated. A
similar analysis of the high frequency band shows the same general trends.
This is clearly an aspect of the data that must be explained by any detailed
model for the behavior discussed here, and such a model is outside the scope
of this paper, as it demands calculation of the effects of aberration and
projection for particular possible flow patterns.

\subsection{Interpretation: Comparison With a Random Walk Model}
 It has been known since observations of polarization rotations in the UMRAO
data and subsequent interpretation by Jones et al. \markcite{REF23} (1985) in
terms of a random walk model, that the turbulent character of the magnetic
field in sources such as BL~Lacs plays a major role in determining their
polarization behavior. In this view, the source is composed of a number of
randomly aligned magnetic field elements, the polarized emissions from which
add to yield a small but non-zero net polarized flux. The flow advects these
`cells' of magnetic field, so that new cells enter through the $\tau=1$
surface, as others are lost to view in the adiabatically-expanded outer
regions of the flow. Thus, at close times, a subset of cells have changed,
but a substantial number are common to the two epochs, leading to a coherence
in the stochastically varying net polarization.

 Rotations arise because once at an extremum in the $Q$-$U$ plane, the
subsequent evolution must by definition have some component towards the
origin, and the coherence of the process guarantees that once a source
evolves in that direction, it will continue to do so for some further time.
In general, the $Q$-$U$ plane locus will pass some distance from the origin,
and eventually do so again at some other polarization angle, having
effectively rotated in $Q$-$U$. Can this effect explain the evolution of the
periodic part of the polarized flux just described? We have constructed a
simple model, in which unit vectors are randomly aligned, and a net polarized
flux was computed assuming that each cell contributes the same absolute
value. Our model employs 100 cells, 5 of which are replaced by new ones at
each of 750 time steps. The resultant behavior of $U$ is superficially
similar to that seen in OJ~287; however, this is misleading.

 Figure~15 displays the evolution of a portion of the $Q$ and $U$ signals,
reconstructed from a wavelet transform of the mock signals. The smoothness
results from the filtering -- the original $Q$-$U$ plane evolution is much
more erratic. A particular direction in this plane appears to have been
picked out, but that arises from the modest number of excursions in the
plane: small number statistics cause a chance alignment of these loops. In
particular, the excursions are randomly aligned with time, there being no
tendency to repeatedly move along a specific orientation, {\it unlike the
variations seen in OJ~287, where during the distinct activity of the 1980s,
the $Q$-$U$ plane excursions are consistently in the same direction}.  We
conclude that such a random walk model cannot explain the character of the
modulated polarization behavior seen in the OJ~287 data.

\section{Discussion}
 We have found that a continuous wavelet transform of the multifrequency,
total flux and polarization data for the BL~Lac object OJ~287 clearly reveals
a periodicity that is merely hinted at by a more conventional Fourier
analysis.  The modulation of the total flux persists for the duration of the
time series available, but changes its period in the 1980s in a way that is
consistent with the kinematics of a `shock-in-jet' model, wherein a portion
of a previously quiescent flow is replaced by a propagating domain of
enhanced emissivity. This may be a single event, or a series of
closely-spaced individual events. In either case, the modulation appears to
be an additional phenomenon, superposed on longer-term trends, and may be a
consequence of a precessional-like motion of the flow in the vicinity of the
$\tau=1$ surface, as a consequence of propagating modes of the
Kelvin-Helmholtz instability.  Such behavior is evident also in the polarized
flux, for which the most notable feature is a series of periodic excursions
along a well-defined axis in the $Q$-$U$ plane which is at about $45^{\circ}$
from the direction defined by the VLBI jet. This behavior differs in detail
from what would be expected from a `random walk' model, and it also is
suggestive of a periodic variation in the direction of the jet, and its mean
magnetic field component, at that point along the flow from which most of the
emission originates in the cm-waveband.

 Detailed modelling is called for to explore this behavior, but will require
that allowance be made for projection and relativistic aberration effects. As
the latter will be determined by the details of the flow's velocity field, it
will be interesting to see if the observed behavior can be a generic feature
of unstable flows, or requires finely-tuned geometry, magnetic field topology
and velocity distribution.

 Were OJ~287 to be in an radio-active phase, intensive, space-based
monitoring of the evolution of VLBI components recently ejected from the core
would provide potentially exciting data, because such a component would
define the detailed kinematics of the flow that dominates the total flux
data. Unfortunately, as evidenced by UMRAO monitoring, the source is in a
protracted quiescent phase, and further understanding of this object may have
to await renewed activity.

 A possibility worth exploration is that OJ~287 is not unique amongst
well-observed sources, but merely a source with particularly short time scale
periodic components. It may well be that some long term variations seen in
other sources are manifestations of a similar modulation, but expressed in
those sources in a far less evident and accessible way: the UMRAO database is
not temporally long enough to easily identify such behavior. We plan to
explore other sources of the UMRAO database, in an attempt to identify
similar activity through signatures such a phase coherence of outbursts over
long times scales.

\acknowledgments
 This research was supported in part by grants AST-9421979 and AST-9617032
from the National Science Foundation. We wish to thank Denise Gabuzda for
comments that helped us improve the presentation of this material.

\clearpage

\figcaption{30 day averaged data for OJ~287. The bottom panel is total flux,
the middle panel polarized flux, and the top panel is the position angle of
the polarization vector ($E$-field). Triangles denote 4.8~GHz data, circles
8.0~GHz data, and crosses 14.5~GHz data.}

\figcaption{Morlet transforms of a.~Gaussian white noise; b.~a sinusoid in
which the frequency halves at the mid-point of the time series. The top panel
is the signal, the second panel is the real part of the transform
coefficients, the third panel is the coefficient modulus, and the fourth
panel is the phase. There is a one-to-one correspondence between epochs in
the time series and those in the transform space; dilation increases from top
to bottom in each panel. Panels two through four are color-coded with the
minimum value shown as blue, and maximum value shown as red.}

\figcaption{A Morlet transform for 14.5~GHz total flux data for the source
OJ~287. The panels are as described for Figure~2. The two `bands' discussed
in the text are indicated by vertical bars to the right of the panel
displaying the real part of the transform.}

\figcaption{Morlet transforms of a.~BL~Lac at 14.5~GHz; b.~NRAO~140 at
14.5~GHz. The panels are as described for Figure~2, although the phase is
not shown.}

\figcaption{The reconstructed signal from the transform of 14.5~GHz data for
OJ~287. The upper panel is for the band of dilations that picks out the
shorter period activity most prominent in the 1980s, while the lower panel
picks out the persistent $1.66$ yr activity.}

\figcaption{The longer period signal from Figure~5 (lower panel), shown as a 
solid line, with the second half of this signal folded back on itself to
show the phase coherence (dashed line) and a sinusoid of period $1.66$ yr
superposed (dotted line).}

\figcaption{The sum (solid line) of the two reconstructed signals (Figure~5) for
14.5~GHz OJ~287 data, added to a tapered boxcar average of the original data
(dotted line).}

\figcaption{The 4.8~GHz--14.5~GHz spectral index for the two reconstructed
signals (Figure~5) for 14.5~GHz OJ~287 data: longer period signal (dotted);
shorter period signal (dashed).}

\figcaption{A Morlet transform for the `shock-in-jet' model signal. The panels
are as described for Figure~2, although the phase is not shown.}

\figcaption{A Morlet transform for 14.5~GHz $U$ data for the source OJ~287. The
top panel is the signal, the second panel is the real part of the transform
coefficients, the third panel is the coefficient modulus. Panels two and
three are color-coded with the minimum value shown as blue, and maximum value
shown as red.}

\figcaption{Evolution in the $Q$-$U$ plane of the periodic part of the 14.5~GHz
OJ~287 polarized flux.}

\figcaption{As for Figure~11, but with a tapered boxcar average added to give a 
more instructive view of $Q$-$U$ plane evolution.}

\figcaption{The cross-correlation function for $I$ and $U$ in the `low
frequency' band of the wavelet transform of OJ~287 at 14.5~GHz.}

\figcaption{The product of the $I$ and $U$ `low frequency' band variations from
the wavelet transform of OJ~287 at 14.5~GHz. The line near to the top of the
plot marks times when the two signals have the same sign.} 

\figcaption{$Q$-$U$ plane evolution for the filtered random walk model.}

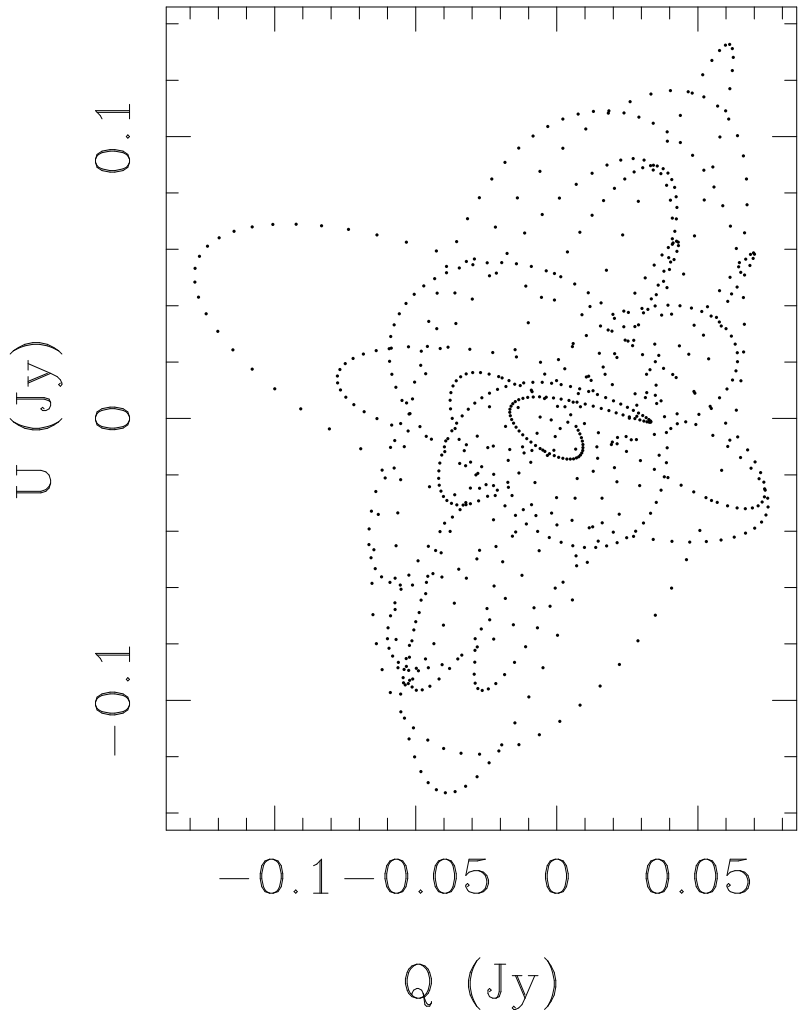
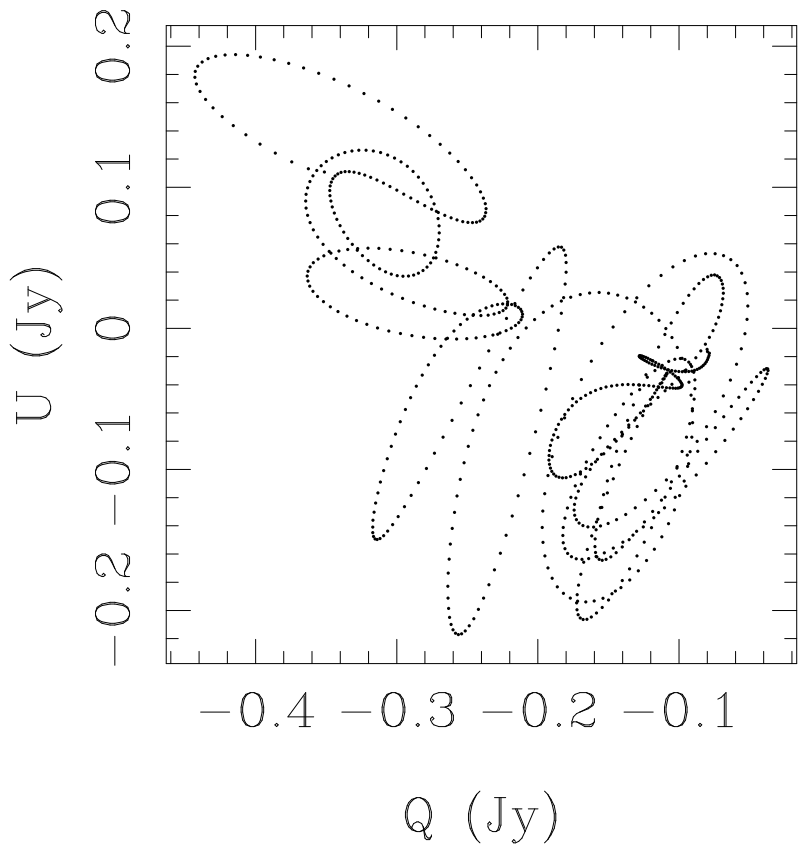
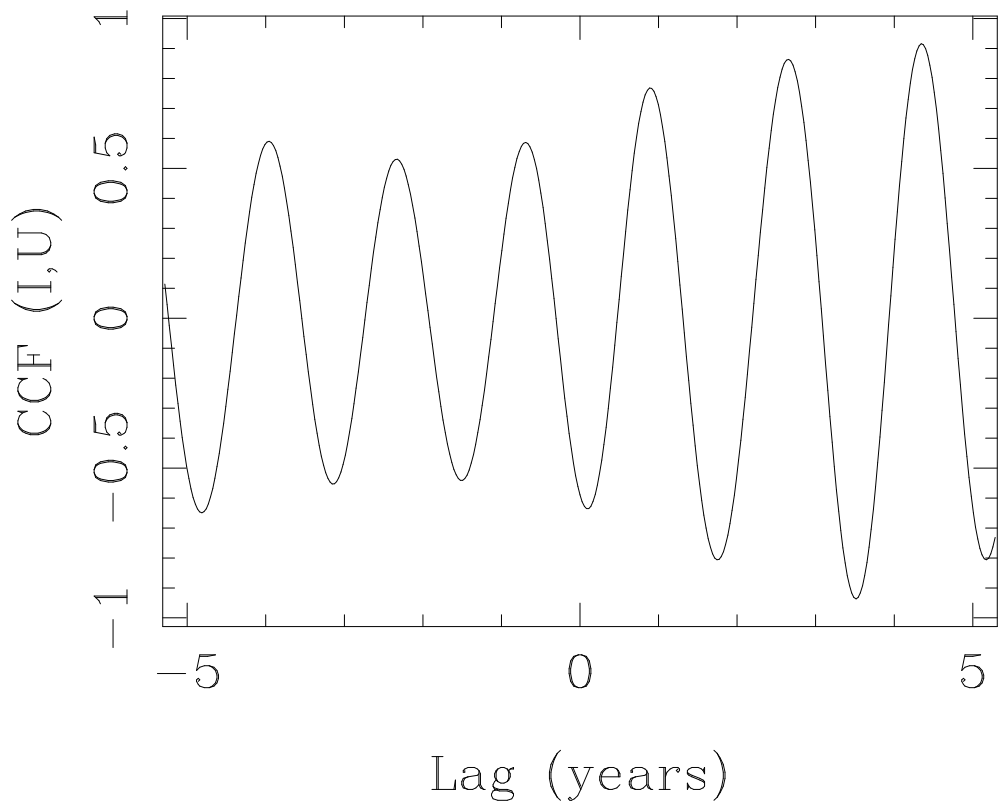
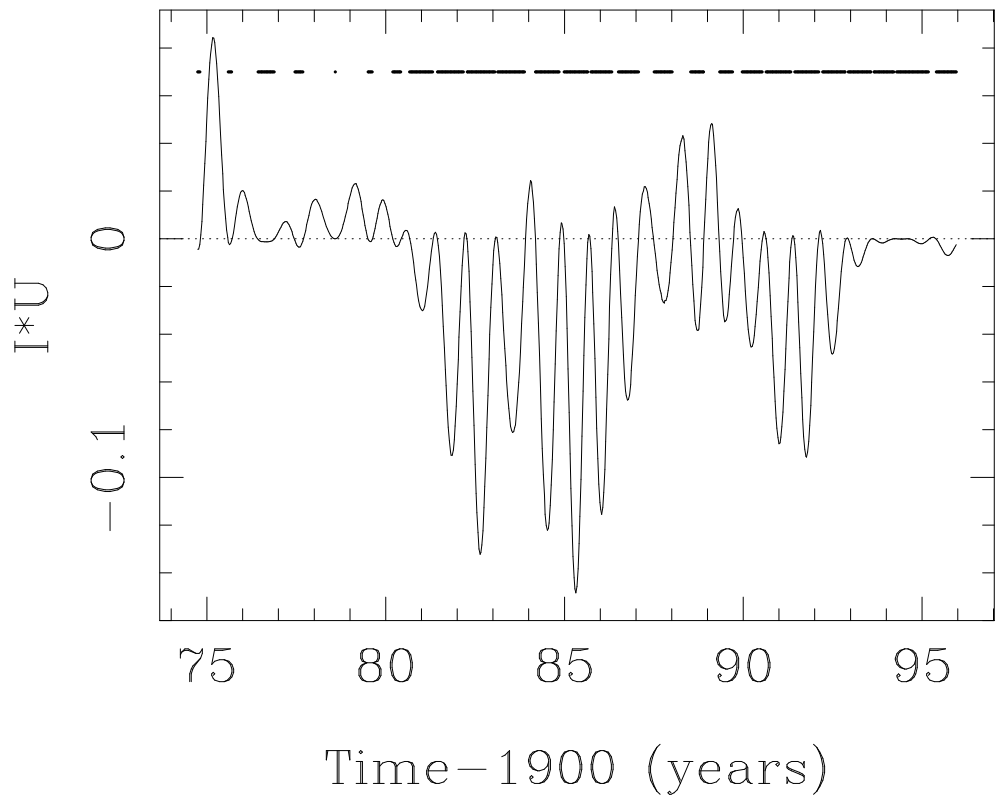
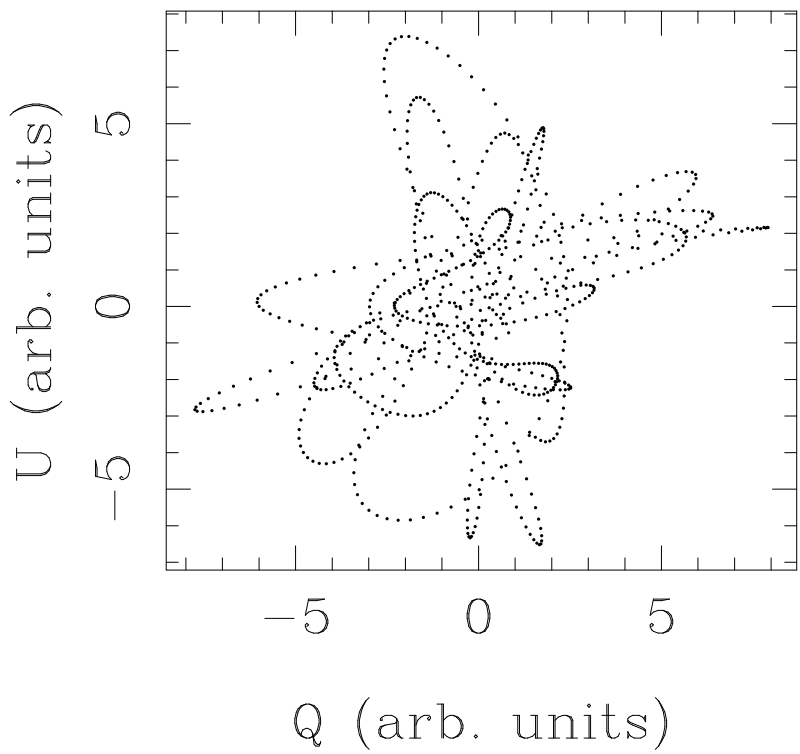
\end{document}